\documentclass[aps,pra,pifont,citesort,twocolumn]{revtex4}
\usepackage{lineno}
\usepackage{color,soul}
 \usepackage{setspace}
\usepackage[utf8]{inputenc}
\usepackage{graphicx,amsmath,amssymb}% Include figure files
\usepackage{dcolumn}% Align table columns on decimal point
\usepackage{bm}% bold math
\usepackage{times}
\usepackage{calligra}
\usepackage{scalefnt}
\usepackage[mathscr]{euscript}

\newcommand{\bk}[1]{\left ( #1\right )}

\newcommand{\eqn}[1]{\begin{eqnarray} \newline #1 \end{eqnarray}}
\newcommand{\ee}{&=&}
\newcommand{\num}[1]{\begin{enumerate} #1 \end{enumerate}}
\newcommand{\hs}{\hspace{0.2cm}}

\newcommand{\bra}[1]{\left \langle#1 \right |}
\newcommand{\ket}[1]{\left |#1\right \rangle}

\newcommand{\nn}{\nonumber}
\newcommand{\half}{\frac{1}{2}}

\newcommand{\com}[2]{\left [ #1,#2 \right]}

\newcommand{\hmin}{H_{\mathrm{min}}}
\newcommand{\hmax}{H_{\mathrm{max}}}

\definecolor{darkgreen}{rgb}{0.0, 0.42, 0.24}

\usepackage{epstopdf} 
\usepackage{epsfig}
\newtheorem{Definition}{Definition}
\newtheorem{Lemma}{Lemma}
\newtheorem{Theorem}{Theorem}
\begin{document}

\title{Composably secure time-frequency quantum key distribution}
\author{Nathan Walk$^{1}$}
\email{nathan.walk@cs.ox.ac.uk}
\author{Jonathan Barrett$^{1}$ and Joshua Nunn$^{2}$}

%\author{Josh}

\affiliation{${}^{1}$Department of Computer Science, University of Oxford, Wolfson Building, Parks Road, Oxford OX1 3QD, United Kingdom\\
${}^{2}$Clarendon Laboratory, University of Oxford, Oxford OX1 3PU, United Kingdom}

\date{\today}

\begin{abstract}
We present a composable security proof, valid against arbitrary attacks and including finite-size effects, for a high dimensional time-frequency quantum key distribution (TFQKD) protocol based upon spectrally entangled photons. Previous works have focused on TFQKD schemes as they combines the impressive loss tolerance of single-photon QKD with the large alphabets of continuous variable (CV) schemes, which enable the potential for more than one bit of secret key per transmission. However, the finite-size security of such schemes has only been proven under the assumption of collective Gaussian attacks. Here, by combining recent advances in entropic uncertainty relations for CVQKD with decoy state analysis, we derive a composable security proof that predicts key rates on the order of Mbits/s over metropolitan distances (40km or less) and maximum transmission distances of up to 140km.

\end{abstract}
\maketitle

\section{Introduction}
Arguably the most promising short term application of quantum information technology is in the field of cryptography, with quantum key distribution (QKD) the canonical example \cite{bb84-orig,Ekert:1991p460}. In the years since its inception, researchers have worked to improve the rigour and generality of security proofs, design protocols that maximise performance and bridge the gap between theoretical proposal and experimental implementation \cite{Lo:2014ex,Scarani:2009p378}. On the security side, one looks to derive a security proof that is {\it composably} secure against arbitrary eavesdropping attacks whilst including all finite-size statistical effects \cite{Renner:2005p464} (see also \cite{Tomamichel:2012p7120}). Practically, one searches for schemes that maximise both the raw clock-rate (the number of transmissions per second) and the number of secure bits per transmission to achieve the largest overall secret key rate at a given transmission distance. 

Most photonic QKD implementations fall into one of two regimes. Traditional discrete variable (DV) schemes encode the secret key in a two-dimensional Hilbert space such as the polarisation degrees of freedom of a single photon. Extending from the original works \cite{bb84-orig,Ekert:1991p460}, these protocols now enjoy universal security proofs \cite{Tomamichel:2012p7120} that function with reasonably small finite-size data blocks, and converge to the ideal Devetak-Winter rates for collective attacks \cite{Devetak:2005p5086} in the asymptotic limit. Continuous variable (CV) schemes utilise an infinite-dimensional Hilbert space, commonly the quadratures of the optical field \cite{Reid:2000p5545,Grosshans:2002p377}. Whilst the finite range and precision of real-life detectors ensures the key is never perfectly continuous, CVQKD nevertheless has the capability to achieve greater than one bit per transmission. Furthermore, composable, general, finite-size CVQKD security proofs have also appeared, although the present results either require extremely large block sizes \cite{Leverrier:2015he}, or are very sensitive to losses \cite{Furrer:2012p8365,Furrer:2014uy} and fail to converge to the Devetak-Winter rates. 

This behaviour is in large part due to the different way loss manifests itself in DV and CV systems. If a single photon is sent through an extremely lossy channel, it will only be detected with very low probability. However, in the instances where a detection does take place, the quantum state is largely preserved and the security is unaffected. Therefore, one can in principle achieve high rates over lossy channels by improving the repetition rate of the photon source or multiplexing. But for coherent or squeezed states commonly used in CVQKD, the loss degrades the signal for all transmissions, rendering the information advantage so small that even modest experimental imperfections will eventually prohibit key extraction.

An alternative approach is to encode the key in the continuous degrees of freedom of single photons, inheriting both the loss tolerance of DVQKD and the larger encoding space of CV protocols \cite{Zhang:2008jh}. These time-frequency schemes are primarily pursued via the temporal and spectral correlations of single photons emitted during spontaneous parametric down conversion (SPDC) and the security stems from the conjugate nature of frequency and arrival time measurements. One can use fast time-resolving detectors to directly measure photon arrival times and a grating spectrometer to measure frequency. It is also possible to adopt just the former detection scheme and convert to frequency measurements via dispersive optics \cite{Mower:2013tu}, or the solely the latter and convert to time via phase modulation \cite{Nunn:2013kf}. Significant progress has been made on the theoretical \cite{Zhang:2014gt} and experimental front \cite{Lee:2014vm,Zhong:iu} however, a general composable security proof is lacking. Exploiting techniques from traditional CVQKD \cite{Navascues:2006p805,GarciaPatron:2006p381,Leverrier:2010p150}, security proofs have been derived against Gaussian collective attacks and extended to incorporate finite-size effects \cite{Lee:2015db} and decoy-states \cite{Bunandar:2015wx} culminating in a result including both \cite{Bao:jh}.

In this work we present a finite-size, composably secure proof for TFQKD by combining the entropic uncertainty proofs for CVQKD \cite{Furrer:2012p8365} with efficient, finite-size decoy-state analysis \cite{Ma:2005ua,Lim:2014uw} for DVQKD. The resultant proofs allow for high rates of key to be distributed over urban and inter-city distances with reasonable block sizes. 

\section{Security Proof I}

\subsection{Generic protocol}
A fairly generic TFQKD decoy-state protocol can be summarised as follows.
\num{\item Quantum transmission and measurement: Quantum states are distributed from Alice to Bob through a potentially eavesdropper controlled quantum channel. In particular, using a pulsed SPDC source she prepares time-frequency entangled photons. Each round of transmission is defined by a time frame of length $T_f$ which is centred about the peak of each pump pulse. Alice randomly varies her pump power between three values $\mu_1, \mu_2, \mu_3$, according to probabilities $\{ p_{\mu_1},p_{\mu_2},p_{\mu_3}=1-p_{\mu_1}-p_{\mu_2}\}$. Immediately after the channel, we make the worst case assumption which is that Eve completely purifies the shared state, $\rho_{AB}$, such that the overall tripartite state, $\ket{ABE}$, is pure. Alice and Bob then randomly switch between measuring the frequency or arrival time of the photons. They choose either the time or frequency measurement for key generation and use the other to check for an eavesdroppers presence. To analyse both possibilities, we will write the two incompatible observables as positive operator valued measurements (POVMs) $(\mathbb{X_A},\mathbb{P_A}$) for Alice and $(\mathbb{X_B},\mathbb{P_B}$) for Bob. Here we will always denote $\mathbb{X}$ as the key generating observable and $\mathbb{P}$ as the check. 

\item Parameter Estimation: Alice and Bob first announce their measurement choices in each round over a public, but authenticated, classical channel and discard all instances where they differ, as well as any instances where two or more detections occur in the same frame. This results in raw, correlated variables $(X_A,X_B)$ which take values $x_A = [x_A^1,x_A^2...x_A^{n_X}],x_B = [x_B^1,x_B^2...x_B^{n_P}]$ which are strings of length $n_X-$, distributed according to a probability distribution $p_{x_A,x_B} = \mathrm{Pr}(X_A = x_A, X_B = x_B)$ and similarly for $P_A$ and $P_B$. Throughout, we will use uppercase to denote random variables and lowercase to denote a corresponding string that is an instantiation of that variable.  %Due to the fundamentally indefinite nature of photon creation in SPDC, as well as technical noise, the measurement strings are actually made up of detections resulting from vacuum, single photon, and multi-photon emissions. In order to apply a non-trivial entropic uncertainty relation, Alice and Bob would like to consider only the single photon subspace. 
Alice then announces which intensity was used in each transmission and the results are further partitioned into substrings e.g. $x_A$ is partitioned into $x_{A,\mu_k}$ of length $n_{X,\mu_k}$ for $k\in \{1,2,3\}$ and similarly for the other strings. Using the number of detections for each pump power and decoy state analysis, Alice and Bob lower bound the number of signals that originated from a single photon transmission. They then announce all outcomes for the $\mathbb{P}$ observables and evaluate the quality of their correlations.
If the quality is sufficiently high (in a way we will make precise later) they proceed, otherwise they abort. Call the passing probability $p_{\mathrm{pass}}$. Conditional on passing, they are left with raw keys which are partially correlated between Alice and Bob as well as the eavesdropper. The overall conditional state between Alice, Bob and Eve is a classical-quantum state of the form, 
\eqn{\rho_{X_AX_BE} = \sum_{x_A,x_B} p_{x_A,x_B}\ket{x_A,x_B}\bra{x_B,x_A}\otimes\rho_E^{x_A,x_B} \label{meas}}
\item Reconciliation: Either Alice or Bob is designated the reference partner, which means that their string is designated as the `correct' string. The reference partner then sends information to the other party to correct any errors between the two strings. If the reference partner is Alice, and the reconciliation information flows in same direction as the quantum transmission this is called direct reconciliation (DR). The converse is called reverse reconciliation (RR). Here we will consider the DR case. If the reconciliation is successful, Alice and Bob will now have perfectly correlated strings $x_B = x_A$ which are still partially known to Eve. In fact, Eve will usually have learned some more information about the strings during the reconciliation process. 
The amount of `leaked' information is denoted $l_{\mathrm{EC}}$. There is also an additional loss from a reconciliation check procedure, where Alice announces a further string of size $\log(1/\epsilon_c)$ to ensure the strings are identical except with probability $\epsilon_c$.

\item Privacy Amplification: Alice and Bob now apply a function, $f$, drawn randomly from a family, $\mathcal{F}$, of two-universal hashing functions to their measurement strings giving $\{f(x_A), f(x_B)\} = \{s_A, s_B\}$. The final state is now
\eqn{\rho_{S_AS_BE} = \sum_{s_A,s_B} p_{s_A,s_B}\ket{s_A,s_B}\bra{s_B,s_A}\otimes\rho_E^{S_A,S_B} \label{S}}

This ideally result in strings  of length $l$ which are perfectly correlated, uniformly random, and completely independent of Eve. These are the final secret keys. The goal of a security analysis is to find a lower bound on the number of extractable bits, $l$, for any given protocol. }%The secret key rate is then just $\frac{l}{N}$.}

\subsection{Composable security}
We now formally state the definitions of composable security and a formalism to quantitatively relax from the ideal case \cite{Renner:2005p464,Tomamichel:2012p7120}.
\begin{Definition}
A protocol that outputs a state of the form (\ref{S}) is
\begin{itemize}
\item {\it $\epsilon_c$-correct} if $\mathrm{Pr}[S_A \neq S_B] \leq \epsilon_c$ and {\it correct} if the condition holds for $\epsilon_c = 0$.
\item $\epsilon_s$-secret if 
\eqn{\hs p_{\mathrm{pass}} \half ||\rho_{S_AE} - \tau_{S_A}\otimes\sigma_E|| \leq \epsilon_s \label{sec}}

where $\rho_{S_AE} = \mathrm{tr}_B(\rho_{S_AS_BE})$, $||\cdot||$ is the trace norm and $\tau_{S_A}$ is the uniform (i.e. maximally mixed) state over $S_A$. It is {\it secret} if the condition holds for $\epsilon_s = 0$.
\end{itemize}
The protocol is {\it ideal} if is is both correct and secret and {\it $\epsilon_{\mathrm{sec}}$-secure} if it is $\epsilon_{\mathrm{sec}}$-indistinguishable from an ideal protocol. This means that there is no device or procedure that can distinguish between the actual protocol and an ideal protocol with probability higher than $\epsilon_{\mathrm{sec}}$. If the protocol is $\epsilon_s$-secret and $\epsilon_c$-correct then it is $\epsilon_{\mathrm{sec}}$-secure for any $\epsilon_{\mathrm{sec}}> \epsilon_c + \epsilon_s$.
\end{Definition}

The choice of error reconciliation fixes $\epsilon_c$ so the goal is now to find a method to bound $\epsilon_s$. First, we briefly introduce the entropic quantities appropriate for finite-size analysis. For a random variable $X$ coupled to a quantum system $E$ associated with a Hilbert space $\mathcal{H}_E$ with the joint system described by a classical-quantum state $\rho_{XE} = \sum_x p_x \ket{x}\bra{x} \otimes \rho_E^x$, the conditional min-entropy of $X$ can be defined as the negative logarithm of the optimal probability of successfully guessing $X$ given $E$ \cite{Konig:2009uj}, that is,
\eqn{\hmin(X|E)_{\rho_{XE}} = -\log\bk{\sup_{\{E_x\}} \sum_x p_x \mathrm{tr}\bk{E_x\rho_E^x}}}
where the supremum is taken over all POVMs and the logarithm here and throughout is taken to be base 2. A related quantity is the conditional max-entropy
\eqn{\hmax(X|E)_{\rho_{XE}} = 2\log\bk{\sup_{\sigma_E}\sum_{x} F(p_X\rho_E^x,\sigma_E)}}
where $F(\rho,\sigma) = \mathrm{tr}\bk{|\sqrt{\rho}\sqrt{\sigma}|}$ is the quantum fidelity and the supremum is over all physical states in $\mathcal{H}_E$, that is $S(\mathcal{H}_E) = \{ \sigma_E \in \mathcal{H}_E|\sigma_E\geq0,\mathrm{tr}(\sigma_E) = 1\}$. One can also define smoothed versions of these quantities that consider $\epsilon$-regions in the state space. Concretely we have,
\eqn{\hmin^\epsilon(X|E)_{\rho_{XE}} \ee \sup_{\tilde{\rho}_{XE}} \hmin(X|E)_{\tilde{\rho}_{XE}} \nn\\
\hmax^\epsilon(X|E)_{\rho_{XE}} \ee \inf_{\tilde{\rho}_{XE}} \hmax(X|E)_{\tilde{\rho}_{XE}}}
where the supremum and infimum are taken over all states $\tilde{\rho}_{XE}$ that are $\epsilon$-close in the purified distance, defined as $\mathcal{P}(\rho,\sigma) = \sqrt{1-F^2(\rho,\sigma)}$. We again emphasise that throughout this work we will be considering the classical-quantum states conditioned on the parameter estimation test having been passed. For the rest of this work we will suppress the state subscript in the entropies.

If the guessing probability is low then the variable $X$ must have a high degree of randomness with respect to an observer holding $E$. Intuitively then, we might expect the conditional smooth min-entropy to be related to the number of secret bits extractable from variable $X$ with failure probability $\epsilon$ as described in Definition 1. This intuition is usefully formalised in the Leftover Hash Lemma (with quantum side information) \cite{Tomamichel:2011ci,Berta:2011p8367}.
\begin{Lemma}
\label{leftover}
Let $\rho_{X_AX_BE}$ be a state of the form (\ref{meas}) where $X_A$ is defined over a a discrete-valued and finite alphabet, E is a finite or infinite dimensional system and $R$ is a register containing the classical information learnt by Eve during information reconciliation. If Alice applies a hashing function, drawn at random from a family of two-universal hash functions \footnote{Let $X,S$ be sets of finite cardinality $|S|\leq|X|$. A family of hash functions $\{\mathcal{F}\}$, is a set of functions $f: X\rightarrow S$ such that $\forall f \in \mathcal{F}, (x,x') \in X$, $\mathrm{Pr}[(f(x) = f(x')]\leq \frac{1}{|S|}$} that maps $X_A$ to $S_A$ and generates a string of length $\it{l}$, then
\eqn{\half||\rho_{S_AE} - \tau_{S_A}\otimes\sigma_E|| \leq \sqrt{2^{l - H_{\mathrm{min}}^\epsilon(X_A|ER)-2}}+2\epsilon \label{hash}}
where $H_{\mathrm{min}}^\epsilon(X_A|ER)$ is the conditional smooth min-entropy of the raw measurement data given Eve's quantum system and the information reconciliation leakage.
\end{Lemma}
Comparing (\ref{sec}) and (\ref{hash}) we see that with an appropriate choice of $l$ we can ensure the security condition is met. In particular we see that the smooth min-entropy is a lower bound on the extractable key length. Suppose that we are only able to bound the smooth min-entropy with a certain probability $1- \epsilon_{\mathrm{fail}}$ (in this work this will be due to the use of Hoeffding's bound in the decoy-state analysis). To get a more exact expression notice that if we choose 
\eqn{l \ee \hmin^\epsilon(X_A|ER) + 2 -2 \log \frac{p_{\mathrm{pass}}}{\epsilon_1}} for some $\epsilon_1>0$ then the r.h.s of (\ref{hash}) is $\epsilon_1/p_{\mathrm{pass}}+2\epsilon$. Then, provided \eqn{\epsilon \leq \frac{\epsilon_s' - \epsilon_1}{2p_{\mathrm{pass}}} \label{ese1}} the convexity and boundedness of the trace distance implies we will satisfy (\ref{sec}) for any secrecy parameter $\epsilon_s \geq \epsilon_s'+ \epsilon_{\mathrm{fail}}$. 
Recalling that by assumption Eve learns at most $l_{EC} + \log 1/\epsilon_c$ bits during information reconciliation we have that,
    
\eqn{\hmin^\epsilon(X_A|ER) &\geq& \hmin^\epsilon(X_A|E) - l_{EC} - \log \frac{1}{\epsilon_c}}
Finally since $\log(p_{\mathrm{pass}}) <0$ we have the following result \cite{Tomamichel:2012p7120,Furrer:2012p8365}
\begin{Theorem}
Let $\rho_{X_AE}$ describe the state between Alice and Eve conditioned on the parameter estimation test succeeding such that the Leftover Hash lemma is applicable. For an error correction scheme as defined above we may extract an $\epsilon_c$-correct and $\epsilon_s$-secret key of length
\eqn{l\geq \hmin^\epsilon(X_A|E) - l_{EC} - \log\frac{1}{\epsilon_c\epsilon_1^2} +2 \label{kth1}}
\end{Theorem}

So the problem has essentially condensed to bounding the conditional smooth min-entropy, $\hmin^\epsilon(X_A|E)$. The central idea is to quantify the smooth min-entropy in one observable by observing the statistics of another, incompatible, observable. This is nothing more than a manifestation of Heisenberg's uncertainty principle, which has long underpinned quantum cryptographic protocols. Specifically, this notion is quantitatively expressed via an uncertainty relation for the smooth min- and max-entropies  \cite{Tomamichel:2011p461} and its extension to the infinite dimensional setting in \cite{Berta:2011p8367,Furrer:2014ig}. These relations can be formulated as follows \cite{Tomamichel:2011p396,Furrer:2012p8365}. Let $\rho_{ABC}$ be an $n_X$-mode state shared between Alice, Bob and Charlie and let Alice's measurements be described by POVMs $\mathbb{X}_A$ and $\mathbb{P}_A$ with elements $\{E_{i}\}$ and $\{F_{j}\}$ respectively. Let $X_A$ be the random variable describing the measurement outcome and $\rho_{X_AC}$ be the joint state of the measurement register and system $C$ given that Alice measured $\mathbb{X}_A$ on each of the $n_X$ modes. Further, let $\mathscr{P}_A$ describe the measurement outcome and $\rho_{\mathscr{P}_AB}$ be the joint state of the measurement register and system $B$ given the counterfactual scenario where Alice instead measured $\mathbb{P}_A$ upon each mode. The sum of the corresponding smooth entropies satisfies the relation
\eqn{\hmin^\epsilon(X_A|C) + \hmax^\epsilon(\mathscr{P}_A|B) \geq -n_X \log c \label{eur}} 
where $c = \max_{i,j} ||\sqrt{E_{i}}\sqrt{F_{j}}||_\infty$ quantifies the compatibility of the measurements with $||\cdot ||_\infty$ the operator norm or the largest singular value.

We now turn to our specific measurement setup where we identify the conjugate measurements $\mathbb{X_A}$ and $\mathbb{P_A}$ with time and frequency.

\subsection{Time-frequency measurement uncertainty relation}
Following \cite{Delgado:1997cm,Zhang:2014gt} we describe the arrival time and conjugate frequency detuning measurements by the following operators,

\eqn{\hat{t}_J \ee \int dt \hs t_J \hs  \hat{E}^\dag(t)_J\hat{E}(t)_J \nn\\
\hat{\omega}_J \ee \int \frac{d\omega}{2\pi} \hs \omega_J \hs  \hat{A}^\dag(\omega)_J\hat{A}(\omega)_J }

for $J \in \{A,B\}$. If we restrict the field operators to the Hilbert space spanned by the single photon time or frequency domain states, $\{ \ket{t_J}:-\infty<t<\infty\}$ and $\{ \ket{\omega_J}:-\infty<\omega<\infty\}$, then we have $\hat{E}_J(t) = \ket{0_J}\bra{t_J}$ and $\hat{A}_J(\omega) = \ket{0_J}\bra{\omega_J}$ so that we can write,
\eqn{\hat{t}_J \ee \int dt \hs t \ket{t_J}\bra{t_J} \nn\\
\hat{\omega}_J \ee \int \frac{d\omega}{2\pi} \hs \omega \ket{\omega_J}\bra{\omega_J} \label{tfop}}
These operators can be shown to be maximally complementary, self-adjoint projectors describing an arrival time measurement that satisfy $\com{\hat{t}_J}{\hat{\omega_K}} = i \delta_{JK}$, and hence can be considered equivalent to the canonical position and momentum operators \cite{Delgado:1997cm}.

Fortunately, the smooth-min entropy uncertainty relations have recently been extended to allow for observables and eavesdroppers living in infinite dimensional Hilbert spaces \cite{Furrer:2011p8368,Furrer:2012p8365,Furrer:2014ig}. However, only in the instances where Alice's source emitted exactly one photon will the POVM's be restricted as per (\ref{tfop}) and result in a useful uncertainty relation. To this end, let $\mathbb{X}_{A,1}$ be a POVM, defined as the restriction of the POVM $\mathbb{X}_A$ to the single photon subspace such that it is described as per (\ref{tfop}). We can now consider the decomposition of the measurement record into variables describing the single, vacuum and multi-photon components components such that we have $\hmin^\epsilon(X_A|E) = \hmin^{\epsilon}(X_{A,1}X_{A,0}X_{A,m}|E)$. In order to apply the uncertainty relation directly we consider the case where Eve assumed to know the multi-photon and vacuum measurements and is left solely with estimating the single photon components, that is we set $C = X_{A,0}X_{A,m}E$ in (\ref{eur}). The following section explains how to relate $\hmin^\epsilon(X_A|X_{A,0}X_{A,m}E)$ to $\hmin^\epsilon(X_A|E)$ and also how to estimate the number of single photon events in a given set of detections. Even though Alice never knows in a given run how many photons are emitted, the number of single-photon events in a collection of runs can be bounded via decoy-state analysis which involves using states with known {\it average} photon numbers. For now we turn to computing the overlap for measurements described by (\ref{tfop}).

In fact, Alice and Bob actually measure coarse grained, finite versions of these measurements. This is a practical necessity in ordinary CVQKD (all homodyne measurements have a finite precision and dynamic range) and in this case, measuring precisely an arrival time operator as defined in (\ref{tfop}) would require a detector that has been turned on in the infinite past. Furthermore, a finite alphabet is necessary in order to apply the leftover hash lemma. In standard CVQKD the quadrature observables can usually be treated symmetrically. In this work we must consider the conjugate observables individually, partly because in practice they have different achievable measurement resolutions and partly because they are physically different quantities. For instance, for arrival time measurements the maximum value is equal to the time frame duration for each measurement round, which in turn puts immediate limits on the maximum overall clock rate of the protocol.

Alice's measurements are divided into evenly spaced bins of width $\delta_{X},\delta_P$ up to a maximum value $\pm \Delta_{X},\pm\Delta_P$ such that $M_{X}=2\Delta_{X}/\delta_{X}+1, M_{P}=2 \Delta_{P}/\delta_{P} +1$ are assumed integer alphabet sizes for simplicity. We can write binned observables corresponding to intervals on the real line $I_{1} = (-\infty, -\Delta_X +\delta_X], I_2 = (-\Delta_X+\delta_X, -\Delta_X+2\delta_X]...I_{M_X} = (\Delta_X-\delta_X,\infty]$. The measurement outcome range is then denoted $\mathcal{X} = \{1,2,...,M_X\} \subset \mathbb{Z}$. Thus the POVM elements of $\mathbb{X}_{A,1}$ are projectors in (\ref{tfop}) integrated over the bin intervals,
\eqn{E_i = \int_{I_i} k_A \ket{k_A}\bra{k_A} dk_A \label{quad}, \hs k_A\in \{t_A, \omega_A\}}
and similarly for $\mathbb{P}_{A,1} = \{F_j\}$. Notice that this is something of a problem as the two infinite end intervals of these binned measurements actually have a large overlap. In fact $||\sqrt{E_1}\sqrt{F_1}|| \approx 1$ which would mean that for these particular measurements the RHS of (\ref{eur}) is approximately zero and the relationship becomes useless. 

To avoid this problem, instead consider a second, hypothetical set of discrete measurements $(\tilde{\mathbb{X}}_{A,1}, \tilde{\mathbb{P}}_{A,1})$ which are defined as per (\ref{quad}) but over a new interval set which is simply the infinite collection of intervals, $\{\tilde{I}_i\}_{i\in \mathbb{Z}}$, of width $\delta$, enumerated such that $\tilde{I}_j = I_j \hs  \forall \hspace{1mm}  k\in \mathcal{X}$. For these measurements the maximum overlap is given by \cite{Furrer:2012p8365},

\eqn{c(\delta_X,\delta_P) = \frac{\delta_X\delta_P}{2\pi}S_0^{(1)}\bk{1,\frac{\delta_X\delta_P}{4}}}
where $S_0^{(1)}(\cdot,u)$ is the radial prolate spheroidal wavefunction of the first kind. Thus, for sufficiently small bin sizes, we can always recover a nontrivial value of $c$ and thus a useful uncertainty relation. The idea is that, for a state that mostly lives in the phase space spanned by the region $[-\Delta_X, \Delta_X]$, the classical-quantum states after Alice applies $\tilde{\mathbb{X}}_{A,1}$ and $\mathbb{X}_{A,1}$ will be very close. We will use our knowledge of Alice's state preparation to quantify this `closeness'. In particular, we will assume that for the all states used in the protocol Alice's source produces a tensor product state, and in particular for the $n_X$ states on which $\mathbb{X}_A$ is measured there is some $\sigma_{AB}$ such that $\rho_{AB} = (\sigma_{AB})^{\otimes {n_X}}$. Moreover, our knowledge of Alice's state allows us to lower bound the probability of measuring a value within the range $[-\Delta_X, \Delta_X]$ on any given run such that,
\eqn{\int_{-\Delta_X}^{\Delta_X} \mathrm{tr}\bk{\sigma_{AB} \ket{k_A}\bra{k_A}} dk_A \geq p_{\Delta_X}}
This it turn means that the probability of measuring an absolute value larger than $\Delta_X$ at any point in the whole protocol given the parameter test was passed is $g(p_{\Delta_X},n_X)/p_{\mathrm{pass}}$ where, 
\eqn{g(p_{\Delta_X},n_X) \leq1- p_{\Delta_X}^{n_X}}
and a similar relation holds for the $\mathbb{P}_A$ measurements.

We then finally have a relation between the entropies of the two discretized measurements conditional on a system $C$, namely \cite{Furrer:2012p8365}
\eqn{\hmin^{\epsilon}(X_{A,1}|C)&>&\hmin^{\epsilon-\epsilon'}(\tilde{X}_{A,1}|C) \nn\\
-\hmax^{\epsilon}(\mathscr{P}_{A,1}|C)&>&-\hmax^{\epsilon-\epsilon''}(\tilde{\mathscr{P}}_{A,1}|C)}
where \eqn{\epsilon' = \sqrt{\frac{2g(p_{\Delta_X},n_X)}{p_{\mathrm{pass}}}},\hs\epsilon'' = \sqrt{\frac{2g(p_{\Delta_P},n_X)}{p_{\mathrm{pass}}}} \label{eprime}} (recall that the scripted variable $\mathscr{P}_A$ is denoting the hypothetical situation where $\mathbb{P}_A$ was measured on the $n_X$ key generating modes instead). Putting all this together with the uncertainty relation (\ref{eur}) finally allows us to write,

\eqn{\hmin^{\epsilon}(X_{A,1}|X_{A,0}X_{A,m}E) &\geq& n_{X,1}\log_2\frac{1}{c(\delta_X,\delta_P)} \nn\\
 &-&\hmax^{\epsilon-\epsilon'-\epsilon''}(\mathscr{P}_{A,1}|B) \label{minmax}}
where $n_{X,1}$ is the number of instances where Alice and Bob measured in the same basis and only a single photon was created. 
In reality however, the measurement record will also include contributions from vacuum and multi-photon terms so we will need a way to determine a lower bound on the min-entropy of the whole string, $\hmin^{\epsilon}(X_{A}|E)$ in terms of $\hmin^{\epsilon}(X_{A,1}|E)$ so that we can apply (\ref{minmax}). We will also require a lower bound on $n_{X,1}$ and an upper bound upon $\hmax^{\epsilon-\epsilon'-\epsilon''}(\mathscr{P}_{A,1}|B)$ based upon the correlations in the $n_P$ measurements of $\mathbb{P}$ observables. Fortunately, all of these can be achieved via decoy-state analysis.

\subsection{Decoy state analysis}
We employ the decoy-state analysis of \cite{Lim:2014uw} which we will recapitulate in our notation for completeness. Recalling the decomposition of the measurements into vacuum, single and multi-photon components we have $\hmin^\epsilon(X_A|E) = \hmin^{\epsilon}(X_{A,1}X_{A,0}X_{A,m}|E)$. Applying a generalisation of the chain rule for smooth-entropies \cite{Vitanov:hz} gives,

\eqn{ \hmin^\epsilon(X_{A}|E) &>& \hmin^{\alpha_1}(X_{A,1}|X_{A,0}X_{A,m}E)\nn \\ 
&+& \hmin^{\alpha_3 + 2\alpha_4+\alpha_5}(X_{A,0}X_{A,m}|E)\nn\\ &-& 2\log_2\frac{1}{\alpha_2} - 1 \nn}
for $\epsilon = 2\alpha_1 + \alpha_2 + \alpha_3 + \alpha_4+\alpha_5$ where $\alpha_i>0$ for all $i$.
Applying the same chain rule to the second term on the rhs gives,
\eqn{\hmin^{\alpha_3 + 2\alpha_4+\alpha_5}(X_{A,0}X_{A,m}|E) &>& \hmin^{\alpha_4}(X_{A,m}|E)\nn\\ &+& \hmin^{\alpha_5}(X_{A,0}|E)\nn\\
 &-& 2\log_2\frac{1}{\alpha_3}- 1 \nn\\
 &\geq& n_{X,0}\log_2 M_X\nn\\ &-& 2\log_2\frac{1}{\alpha_3}- 1 \nn}
where $n_{X,0}$ is the number of $X$ basis measurements that resulted when the source produced a vacuum state. In the second inequality we have used that $\hmin^{\alpha_4}(X_{A,m}|E) \geq 0$, which is equivalent to assuming all multi-photon events are insecure and also that $\hmin^{\alpha_5}(X_{A,0}|E)\geq \hmin(X_{A,0}|E) = \hmin(X_{A,0}) = n_{X,0} \log_2M_X$ where the inequality is true by definition and final equality comes from assuming that vacuum contributions are uncorrelated with the chosen bit values and uniformly distributed across the measurement range. Note that since $\alpha_4$ and $\alpha_5$ now no longer feature directly, we can set them arbitrarily small and neglect them from further calculations. Putting this together gives,

\eqn{\hmin^\epsilon(X_A|E)&\geq&\hmin^{\alpha_1}(X_{A,1}|X_{A,0}X_{A,m}E) + n_{X,0}\log_2M_X \nn \\
 &-& \log_2\frac{1}{\alpha_3^2\alpha_2^2} - 2}
which we can now bound according to (\ref{minmax}) to get 
\eqn{\hmin^\epsilon(X_{A}|E)&\geq& n_{X,1}\log_2\frac{1}{c(\delta_X,\delta_P)}\nn\\ &-&\hmax^{\alpha_1-\epsilon'-\epsilon''}(\mathscr{P}_{A,1}|B)
+ n_{X,0}\log_2M_X\nn\\ &-& \log_2\frac{1}{\alpha_3^2\alpha_2^2} - 2 \label{hminbnd}}

Now, we also need to derive lower bounds upon the number of vacuum and single photon contributions. Recall that in the protocol, Alice probabilistically selects a pump power, $\mu_k$, with probability $p_k$ which in turn probabilistically results in an $n$-photon state with conditional probability \eqn{p_{n|\mu_k} = \frac{e^{-\mu_k}\mu_k^n}{n!} \label{pnk}} assuming a Poissonian source. Although we cannot directly know how many detections are due to a particular photon number emission, we do know how many detections are due to a particular pump power. The main idea of a decoy state analysis is to use the latter information to place bounds on the former. Following \cite{Ma:2005ua,Lim:2014uw} we first note from the eavesdropper's perspective it could just as well be a counterfactual scenario where Alice instead creates n-photon states and merely probabilistically partitions them so that each subset has a mean photon number $\mu_k$. Indeed, Bayes' rule allows us to write the down the appropriate probability of pump power given $n$-photon emission as,
\eqn{p_{\mu_k|n} = \frac{p_{\mu_k}p_{n|\mu_k}}{\tau_n} \label{pkn}}
where 
\eqn{\tau_n = \sum_k p_{\mu_k}\frac{e^{-\mu_k}\mu_k^n}{n!}} is the total probability of an $n$-photon emission. Note that technically all of these probabilities should also be conditioned on the parameter test on the $\mathbb{P}_A$ basis measurements passing. However, when considering the $\mathbb{X}_A$ basis Alice can be sure that this conditioning will make no difference. To see this, consider the counterfactual case where she prepares $n$-photon states. By simply not assigning $\mu$ values in the $\mathbb{X}_A$ basis until after the parameter test on the $\mathbb{P}_A$ is completed she can ensure that probabilities like (\ref{pkn}) are unchanged by conditioning.
In the asymptotic limit of large statistics, (\ref{pkn}) allows us to relate the number of coincidences given a certain pump power, $n_{X,\mu_k}$ to the number given an $n$-photon emission, $n_{X,n}$, via
\eqn{n^*_{X,\mu_k} \ee \sum_{n=0}^\infty p_{\mu_k|n}n_{X,n}\nn\\
\ee \sum_{n=0}^\infty \frac{p_{\mu_k}e^{-\mu_k}\mu_k^n}{\tau_n n!} n_{X,n}}
where $n^*_{X,\mu}$ is the asymptotic value of $n_{X,\mu_k}$ and we have substituted in from (\ref{pkn}) and (\ref{pnk}). We can then use Hoeffding's inequality for independent events which says that the difference between observed statistics and their asymptotic values is bounded by
\eqn{|n^*_{X,\mu_k}- n_{X,\mu_k}| \leq \lambda(n_X,\epsilon_2)}
and hence $n^{-}_{X,\mu_k}\leq n^{*}_{X,\mu_k} \leq n^{+}_{X,\mu_k}$ where,
\eqn{n^{\pm}_{X,\mu_k} := n_{X,\mu_k} \pm \lambda(n_X,\epsilon_2)}
with probability at least $1-2\epsilon_2$ where $\lambda(n_X,\epsilon_2) = \sqrt{\frac{n_X}{2} \ln \frac{1}{\epsilon_2}}$.
Now consider the following expression:
\eqn{&&\frac{\mu_2e^{\mu_3}n^*_{X,\mu_3}}{p_{\mu_3}} - \frac{\mu_3e^{\mu_2}n^*_{X,\mu_2}}{p_{\mu_2}}\nn\\
 \ee  \sum_{n=0}^\infty \bk{\frac{\mu_2e^{\mu_3} p_{\mu_3}e^{-\mu_3}\mu_3^n n_{X,n}}{n!\tau_n p_{\mu_3}} - \frac{\mu_3e^{\mu_2} p_{\mu_2}e^{-\mu_2}\mu_2^n n_{X,n}}{n!\tau_n p_{\mu_2}}} \nn\\
 \ee \mu_2\mu_3 \sum_{n=0}^\infty \frac{(\mu_3^{n-1} - \mu_2^{n-1})n_{X,n}}{\tau_n n!}\nn}
Notice that in the above expression the summand vanishes when $n=1$. This means we can split up the sum as,
\eqn{&&\frac{\mu_2e^{\mu_3}n^*_{X,\mu_3}}{p_{\mu_3}} - \frac{\mu_3e^{\mu_2}n^*_{X,\mu_2}}{p_{\mu_2}}\nn\\
\ee \frac{(\mu_2 - \mu_3)n_{X,0}}{\tau_0}- \mu_2\mu_3 \sum_{n=2}^\infty \frac{(\mu_2^{n-1} - \mu_3^{n-1})n_{X,n}}{\tau_n n!}\nn\\
&\leq& \frac{(\mu_2 - \mu_3)n_{X,0}}{\tau_0}}
where the inequality holds provided $\mu_2>\mu_3$. Rearranging gives a lower bound on the vacuum conincidences, 
\eqn{n_{X,0} &\geq& n_{X,0}^-:=\tau_0 \frac{e^{\mu_3}\mu_2n^-_{X,\mu_3} - e^{\mu_2}\mu_3n^+_{X,\mu_2}}{p_{\mu_3}p_{\mu_2}(\mu_2 - \mu_3)}}
which holds with probability at least $1 - 4\epsilon_2$.

The single photon bound is somewhat more involved. First, by similar reasoning as above, we have:
\eqn{&&\frac{e^{\mu_2}n^*_{X,\mu_2}}{p_{\mu_2}} - \frac{e^{\mu_3}n^*_{X,\mu_3}}{p_{\mu_3}}\nn\\
\ee \sum_{n=0}^\infty \frac{(\mu_2^{n} - \mu_3^{n})n_{X,n}}{\tau_n n!} \nn\\
\ee \frac{(\mu_2 - \mu_3)n_{X,1}}{\tau_1}+ \sum_{n=2}^\infty \frac{(\mu_2^{n} - \mu_3^{n})n_{X,n}}{\tau_n n!} \label{s11}}
since now the $n=0$ term vanishes. Now, using the identity $a^n -b^n = (a-b)\sum_{i=0}^{n-1}a^{n-1-i}b^i$ we have
\eqn{\mu_2^{n} - \mu_3^{n} = (\mu_2 - \mu_3) \sum_{i=0}^{n-1}\mu_2^{n-1-i}\mu_3^i}
which combined with the inequality $\sum_{i=0}^{n-1}\mu_2^{n-1-i}\mu_3^i \leq (\mu_2 + \mu_3)^{n-1}$ $\forall n\geq 2$
gives
\eqn{\mu_2^{n} - \mu_3^{n} &\leq& (\mu_2 - \mu_3)(\mu_2 + \mu_3)^{n-1}\nn\\
\ee \frac{\mu_2 - \mu_3}{\mu_2 + \mu_3} (\mu_2 + \mu_3)^{n} \nn\\
\ee  \frac{\mu_2^2 - \mu_3^2}{(\mu_2 + \mu_3)^2} (\mu_2 + \mu_3)^{n} \nn \\ 
&\leq & \frac{\mu_2^2 - \mu_3^2}{\mu_1^2} \mu_1^n}
where the second last equality results in a tighter bound when we apply the condition $\mu_1 > \mu_2 + \mu_3$ to obtain the last inequality. Substituting back in (\ref{s11}) yields:
\eqn{&&\frac{e^{\mu_2}n^*_{X,\mu_2}}{p_{\mu_2}} - \frac{e^{\mu_3}n^*_{X,\mu_3}}{p_{\mu_3}}\nn\\
&\leq& \frac{(\mu_2 - \mu_3)n_{X,1}}{\tau_1}+ \frac{\mu_2^2 - \mu_3^2}{\mu_1^2}\sum_{n=2}^\infty \frac{\mu_1^nn_{X,n}}{\tau_n n!}\label{n1}}
Rewriting the sum as
\eqn{\sum_{n=2}^\infty \frac{\mu_1^nn_{X,n}}{\tau_n n!} \ee \sum_{n=0}^\infty \frac{\mu_1^nn_{X,n}}{\tau_n n!} - \frac{n_{X,0}}{\tau_0} - \frac{\mu_1n_{X,1}}{\tau_1} \nn\\
\ee  \frac{e^{\mu_1}}{p_{\mu_1}}n^*_{X,\mu_1} - \frac{n_{X,0}}{\tau_0} - \frac{\mu_1n_{X,1}}{\tau_1}}
and substituting back into (\ref{n1}), we can solve for $n_{X,1}$, and using the Hoeffding bounds arrive at the following lower bound for the single photon detections:
\eqn{n_{X,1} &\geq& n_{X,1}^- :=\frac{\mu_1 \tau_1}{\mu_1(\mu_2 - \mu_3) - (\mu_2^2 - \mu_3^2)}\left [\frac{e^{\mu_2}}{p_{\mu_2}}n^-_{X,\mu_2} \right . \nn \\
 &-& \left. \frac{e^{\mu_3}}{p_{\mu_3}}n^+_{X,\mu_3} + \frac{\mu_2^2 - \mu_3^2}{\mu_1^2} \bk{\frac{n^-_{X,0}}{\tau_0} - \frac{e^{\mu_1}}{p_{\mu_1}}n^+_{X,\mu_1}} \right ]\label{nx1}}
which holds with probability at least $1-6\epsilon_2$.

Now the only unbounded term in the key rate formula is the max-entropy term $\hmax^{\alpha_1-\epsilon'-\epsilon''}(\mathscr{P}_{A,1}|B)$. Firstly, by the data processing inequality we have $\hmax^{\alpha_1-\epsilon'-\epsilon''}(\mathscr{P}_{A,1}|B) \leq \hmax^{\alpha_1-\epsilon'-\epsilon''}(\mathscr{P}_{A,1}|\mathscr{P}_{B,1})$. We again use the results of \cite{Furrer:2012p8365}, where a statistical bound on the smooth max-entropy over a classical probability distribution is found based on the observed correlations. Alice and Bob quantify the correlations by computing the average distance (essentially the Hamming distance but for non-binary strings) which for two strings $p_A$ and $p_B$  taking values in $\mathbb{R}$ is defined as: 
\eqn{d(p_A,p_B): = \frac{1}{n_P} \sum_{i=1}^{n_P} |p_{A}^i - p_{B}^i|
:= \frac{m_P}{n_P} \label{dpe}}
In order to bound $\hmax^{\alpha_1-\epsilon'-\epsilon''}(\mathscr{P}_{A,1}|\mathscr{P}_{B,1})$ we proceed in three steps. Firstly, we use decoy-state arguments to upper bound $d(p_{A,1},p_{B,1})$, the average distance on just the single photon terms. Then, following \cite{Furrer:2012p8365}, we use this upper bound and a result by Serfling \cite{Serfling:1974dx} to upper bound the average distance that could be observed on the counterfactual variables $d({\scriptstyle\mathscr{P}}_{A,1},{\scriptstyle\mathscr{P}}_{B,1})$. Finally, we use this quantity to upper bound the smooth max-entropy.

The quantity $m_{P}$ in (\ref{dpe}) is just counting the number of bins between Alice and Bob's measurements. Considering the substring corresponding to pump power $\mu_1$, in the asymptotic limit, we expect $m_{P,\mu_1}^*$ from $m_{P}$ errors to be assigned to $\mu_1$ 
where
\eqn{m_{P,\mu_1}^* = \sum_{n=0}^\infty p_{\mu_1|n}m_{P,n}}
and $m_{P,n}$ is the number of errors in the $\mathbb{P}$ basis resulting from $n$-photon states.
Just as when we were bounding the number of single-photon terms, we can use Hoeffding's result to bound the difference between this unknown asymptotic quantity and the observed value,
\eqn{m_{P,\mu_1}^*\leq m_{P,\mu_1}^+ = m_{P,\mu_1}+ \lambda'(\epsilon_1,n_P,M_P)}
except with probability $1-2\epsilon_2$ where now $\lambda'(\epsilon_2,n_P,M_P) = \sqrt{\frac{m_PM_P^2}{2} \ln \frac{1}{\epsilon_2}}$ to account for the non-binary nature of entries in the error strings.
Hence we expect in the asymptotic limit to have 
\eqn{m^*_{P,\mu_k} \ee \sum_{n=0}^{\infty} p_{\mu_k | n} m_{P,n}   \geq p_{\mu_k | 1} m_{P,1}\nn\\
\ee p_{\mu_k | 1} n_{P,1} d(p_{A,1} , p_{B,1})}
%\eqn{d^*(p_{A,\mu_k},p_{B,\mu_k}) \ee \frac{m_{P,\mu_k}^*}{n^*_{P,\mu_k}} \nn \\
%\ee \frac{\sum_{n=0}^\infty p_{\mu|n}m_{P,n}}{n^*_{P,\mu_k}} \nn \\
%\ee \frac{\sum_{n=0}^\infty p_{\mu|n}n_{P,n}d(p_{A,n},p_{B,n})}{n^*_{P,\mu_k}} \nn \\
%&\geq & \frac{p_{\mu_k|1} n_{P,1}d(p_{A,1},p_{B,1})}{n^*_{P,\mu_k}}} 
Rearranging gives,
\eqn{d(p_{A,1} , p_{B,1}) &\leq& \frac{m^*_{P,\mu_k}}{p_{\mu_k|1} n_{P,1} }\nn\\
  &\leq&  \frac{m^{+}_{P,\mu_k}}{p_{\mu_k|1} n^{-}_{P,1} }\nn\\
  &:=& d^{+}_{P,1}}
   
with probability at least $1-4\epsilon_2$ where $n_{P,1}^-$ is calculated in the same manner as (\ref{nx1}). Now, say that Alice and Bob abort the protocol whenever $d_{P,1}^+>d_0$. 

Now, we again consider bounding the counterfactual average distance $d({\scriptstyle\mathscr{P}}_{A,1},{\scriptstyle\mathscr{P}}_{B,1})$. For brevity we define $d_{\mathscr{P},1} = d({\scriptstyle\mathscr{P}}_{A,1},{\scriptstyle\mathscr{P}}_{B,1})$ and $d_{P,1}=d(p_{A,1},p_{B,1}) $ and denote the total average distance that would be observed on the combination of the strings as $d_{P,\mathrm{tot}}$. Given that the observed correlations pass the parameter estimation test, we are interested in the probability that the average distance of the hypothetical measurements would be greater than $d_{P,1}$ by a fixed amount.
\eqn{\mathrm{Pr}[d_{\mathscr{P},1}> d^+_{P,1} + C|``\mathrm{pass}"]&\leq& \mathrm{Pr}[d_{\mathscr{P},1}> d_{P,1} + C|``\mathrm{pass}"]\nn \\
&\leq& \frac{\mathrm{Pr}[d_{\mathscr{P},1}> d_{P,1} + C]}{p_\mathrm{pass}}\label{pass}}
where we have used Bayes' theorem in the last line.

Bounding $\mathrm{Pr}[d_{\mathscr{P},1}> d_{P,1} + C]$ is a standard problem of random sampling without replacement.  Defining the total number of detections coming from single photons as $N_1 = n_{X,1} + n_{P,1}$ we have,
\eqn{N_1 d_{P,\mathrm{tot}} = n_{X,1}d_{\mathscr{P},1} + n_{P,1}d_{P,1} \label{drel}}
A result by Serfling shows that for any $a$ \cite{Serfling:1974dx},
\eqn{\mathrm{Pr}[d_{\mathscr{P},1}> a + C|d_{P,\mathrm{tot}} = a] \leq \exp\bk{\frac{-2 n_{X,1}N_{1} C^2}{(n_{P,1}+1) M_P^2}}\label{serf}}
where we recall that $M_{P}$ is the size of the alphabet of $\mathbb{P}_A$ outcomes. Now using (\ref{drel}) and (\ref{serf}) we can write,
\eqn{&&\mathrm{Pr}[d_{\mathscr{P},1}> d_{P,1} + C] = \mathrm{Pr}[d_{\mathscr{P},1}> d_{P,\mathrm{tot}} + \frac{n_{P,1}}{N_1}C]\nn\\
\ee \sum_a \mathrm{Pr}[d_{P,\mathrm{tot}}=a]\mathrm{Pr}[d_{\mathscr{P},1}> a + \frac{n_{P,1}}{N_1}C|d_{P,\mathrm{tot}} = a]\nn \\
&\leq& \exp\bk{\frac{-2 n_{X,1} (n_{P,1})^2C^2}{(n_{P,1}+1)N_{1} M_P^2}}}
Substituting back into (\ref{pass}) and recalling that the protocol aborts whenever $d_{P,1}^+>d_0$ we have,
\eqn{\mathrm{Pr}[d_{\mathscr{P},1}> d_0 + C|``\mathrm{pass}"] \leq\frac{\exp\bk{\frac{-2 n_{X,1}^- (n_{P,1}^-)^2C^2}{(n_{P,1}^++1)N_{1}^+ M_P^2}} }{p_{\mathrm{pass}}} \label{expbound}}
where we have substituted in the lower bounds in the numerator and upper bounds in the denominator. In order to evaluate (\ref{expbound}) we still require the upper bound $N_1^+$, noting that this will automatically yield $n_{P,1}^+ = N_1^+ - n_{X,1}^-$. To this end, define $n_{\mu_k}$ as the total number of detections in both bases at a given pump power, $n_{\mu_k}^*$ as its asymptotic value and  $N_n$ as the number of detections from $n$-photon states. Then we may write,
\eqn{\frac{e^{\mu_2} n^*_{\mu_2}}{p_{\mu_2}} - \frac{e^{\mu_3} n^*_{\mu_3}}{p_{\mu_3}} \ee \sum_{n=0}^\infty \frac{(\mu_2^n - \mu_3^n)N_n}{\tau_n n!} \nn\\
\ee \frac{(\mu_2 - \mu_3)N_1}{\tau_1} + \sum_{n=2}^\infty \frac{(\mu_2^n - \mu_3^n)N_n}{\tau_n n!} \nn \\
&\geq& \frac{(\mu_2 - \mu_3)N_1}{\tau_1}}
provided $\mu_2>\mu_3$ which implies
\eqn{N_1 \leq N_1^+:= \frac{\tau_1}{\mu_2 - \mu_3}\bk{\frac{e^{\mu_2} n^+_{\mu_2}}{p_{\mu_2}} - \frac{e^{\mu_3} n^-_{\mu_3}}{p_{\mu_3}}}}
except with probability $1-4\epsilon_2$.
Finally, we use the following result \cite{Furrer:2012p8365} (Proposition 1),
\begin{Lemma}
Let $\mathcal{P}$ be a finite alphabet, $Q(p,p')$ a probability distribution on $\mathcal{P}^n\times \mathcal{P}^n$ for some $n\in \mathbb{N}$, $\kappa>0$ and $\nu>0$. If $\mathrm{Pr}_{Q}[d(p,p')\geq \kappa] \leq \nu^2$ then,
\eqn{\hmax^\nu(P|P')< n\log \gamma(\kappa)}
where
\eqn{\gamma(x) = (x + \sqrt{1+x^2})\bk{\frac{x}{\sqrt{1+x^2}-1}}^x}
\end{Lemma}
This result might seem surprising given that an entropy is by definition label-independent, whereas the average distance explicitly depends upon the choice of labels. The resolution is that the lemma is derived by taking a worse case scenario in which the number of observed bin errors is assumed to be due to individual entries each of which different by only one bin, thus maximising the max-entropy. This means that the bound will hold true regardless of the labelling convention used on the data, but a poor choice of labelling (for instance one that numbered adjacent bins by greatly differing numbers) would result in a very pessimistic bound. We can apply this result by setting $\nu^2  = \exp\bk{\frac{-2 n_{X,1}^- (n_{P,1}^-)^2C^2}{(n_{P,1}^++1)N_{1}^+ M_P^2}}/ p_{\mathrm{pass}}$. This allows us to bound
\eqn{\hmax^{\nu}(\mathscr{P}_{A,1}|\mathscr{P}_{B,1}) \leq \log_2 \gamma(d_0 +C) \label{maxbound}}
where
\eqn{C \ee M_P \sqrt{\frac{N_1^+(n_{P,1}^++1)}{n_{X,1}^-(n_{P,1}^-)^2} }\nn \\
&\times& \sqrt{\ln \frac{1}{\sqrt{p_{\mathrm{pass}}}\nu}}\label{C1}}
for any $p_{\mathrm{pass}}$ which is always possible provided $\nu>0$. Thus, provided $\alpha_1 - \epsilon' - \epsilon'' >0$ there is some $C$ such that we can set $\nu = \alpha_1 - \epsilon' - \epsilon''$ and use this result to bound the smooth max-entropy in (\ref{hminbnd}). 

The final step is to account for all the error terms due to finite-size effects to find the actual secrecy parameter and to eliminate the explicit dependence upon $p_{\mathrm{pass}}$. From the decoy state analysis we can rewrite $\alpha_1 = (\epsilon - \alpha_2 - \alpha_3)/2$ (recall that we can neglect the $\alpha_4$ and $\alpha_5$ terms). From our security definitions, provided (\ref{ese1}) is satisfied, we will extract an $\epsilon_s = \epsilon_s' + \epsilon_{\mathrm{fail}}$ secret key. In particular we may satisfy (\ref{ese1}) by choosing $\epsilon = \frac{\epsilon_s' - \epsilon_1}{2 \sqrt{p_{\mathrm{pass}}}}$ in which case we have,
\eqn{\sqrt{p_{\mathrm{pass}}}\nu \ee \frac{1}{2} \bk{\frac{\epsilon'_s - \epsilon_1}{2} - \sqrt{p_{\mathrm{pass}}}(\alpha_2 + \alpha_3) } \nn \\
 &-&  \sqrt{(2g(p_{\Delta_X},n_X))} - \sqrt{(2g(p_{\Delta_P},n_X))}  \nn \\
 &\geq& \frac{1}{2} \bk{\frac{\epsilon'_s - \epsilon_1}{2} - (\alpha_2 + \alpha_3) } \nn \\
 &-& \sqrt{(2g(p_{\Delta_X},n_X))} - \sqrt{(2g(p_{\Delta_P},n_X))} \nn\\
 \ee \frac{1}{2} \bk{\frac{\epsilon_s - \epsilon_{\mathrm{fail}} - \epsilon_1}{2} - (\alpha_2 + \alpha_3) } \nn \\
 &-& \sqrt{(2g(p_{\Delta_X},n_X))} - \sqrt{(2g(p_{\Delta_P},n_X))} \label{nu}}
where the second line used $p_{\mathrm{pass}} \leq1$. This lower bound on $\nu$ can be used to upper bound on the logarithmic term in (\ref{C1}).

We must include the failure probabilities from the Hoeffding bounds, which we applied to the number of counts for three pump powers in two measurement bases, each contributing an error term $2\epsilon_2$. This gives an overall error budget $\epsilon_{\mathrm{fail}} = 12\epsilon_2$. If, for simplicity, we choose $\epsilon_{2} = \alpha_2 = \alpha_3 := \epsilon_1$ and set $\epsilon_1 = \epsilon_{s}/21$ then straightforward substitution into (\ref{nu}), which is used to bound (\ref{C1}) and hence  (\ref{maxbound})and  (\ref{hminbnd}) gives us a final expression for the $\epsilon_c$-correct, $\epsilon_{s}$-secret key length:

\eqn{l&\geq&  -n^{-}_{X,1}\log_2c(\delta_X,\delta_P) -n^{-}_{X,1}\log_2\gamma(d_0 + C')\nn\\
 &-& 4\log_2\frac{21}{\epsilon_s} + n_{X,0}\log_2M_X - l_{EC} - \log_2\frac{1}{\epsilon_c\epsilon_s} \label{kfinal}}
where 
\eqn{C' \ee M_P \sqrt{\frac{N_1^+(n_{P,1}^++1)}{n_{X,1}^-(n_{P,1}^-)^2} }\nn \\
&\times& \sqrt{\ln \frac{1}{\epsilon_s/21 - \sqrt{(2g(p_{\Delta_X},n_X))} - \sqrt{(2g(p_{\Delta_P},n_X))}  }}  \nn}
As noted earlier, in order for there to be a positive keyrate, the denominator inside the logarithmic term in $C$ must positive. This means that $\Delta_X,\Delta_P$ are not free parameters, but must be chosen to ensure that this condition is satisfied.

\section{Numerical Evaluation}
We now turn to the the numerical evaluation of the key rate formula, taking parameters mostly from \cite{Lee:2014vm,Zhong:iu} for the simulations. We will consider transmission through optical fibre at telecom wavelengths, which is well modelled as a lossy channel where the transmission is related to the distance, $L$, via $T = 10^{-0.02L}$. When the number of channel uses $N$ (instances where Alice attempts to generate a pair of photons and transmit one to Bob) is large and each party chooses to measure the $\mathbb{X}(\mathbb{P})$ observable with probability $p_X (1-p_X)$ the number of observed counts after sifting for a given pump power will be well approximated by $n_{X,\mu_k} = p_X^2p_{\mu_k}\kappa_{\mu_k}N (n_{P,\mu_k} = (1-p_X)^2p_{\mu_k}\kappa_{\mu_k}N)$ where $\kappa_{\mu_k}$ is the coincidence probability of at least one photon being detected by both Alice and Bob. It is given by \cite{Bunandar:2015wx},
\eqn{\kappa_{\mu_k} &=& \sum_{n = 0}^\infty p_{n|\mu_k} (1 - (1-p_d)(1-\eta_A)^n)\nn\\
&\times& (1 - (1-p_d)(1-\eta_B T)^n) }
where $p_d$ are the dark count probabilities for Alice and Bob's detectors and $\eta_A$ and $\eta_B$ are their respective efficiencies.

We consider an SPDC source that generates a photon pair with temporal wave function
\eqn{\ket{\Psi_{AB}} = \int dt_A dt_B \hs e^{i\omega_P(t_A+t_B)/2}\psi(t_A,t_B)\ket{t_A,t_B}}
where 
\eqn{\psi(t_A,t_B) = \frac{\exp\bk{\frac{-(t_A-t_B)^2}{4\sigma_{\mathrm{cor}}^2} -\frac{(t_A+t_B)^2}{16\sigma_{\mathrm{coh}}^2}}}{\sqrt{2 \pi\sigma_{\mathrm{cor}}\sigma_{\mathrm{coh}}}}\nn}
and $\sigma_{\mathrm{coh}}$ and $\sigma_{\mathrm{cor}}$ are the pump coherence  and photon correlation times respectively. The variance and covariance of Alice and Bob's measurement strings will be
\eqn{V_{t_A} \ee V_{t_B} = \nn \sigma_{\mathrm{coh}}^2 + \frac{\sigma_{\mathrm{cor}}^2}{4}\\
\left < t_A t_B \right > \ee \sigma_{\mathrm{coh}}^2 - \frac{\sigma_{\mathrm{cor}}^2}{4}}

One can also write this as a spectral wave function,
\eqn{\psi(\omega_A,\omega_B) = \frac{\exp\bk{\frac{-\sigma_{\mathrm{cor}}^2(\omega_A-\omega_B)^2}{4} -\sigma_{\mathrm{coh}}^2(\omega_A+\omega_B)^2}}{\sqrt{2 \pi\sigma_{\mathrm{cor}}\sigma_{\mathrm{coh}}}}\nn}
with spectral variances and correlations,
\eqn{V_{\omega_A} \ee V_{\omega_B} = \nn \frac{1}{16} \left(\frac{1}{\sigma_{\mathrm{coh}}^2}+\frac{4}{\sigma_{\mathrm{cor}}^2}\right)\\
\left < \omega_A \omega_B \right > \ee \frac{-\sigma _{\text{cor}}^2+4 \sigma _{\text{coh}}^2}{16 \sigma _{\text{coh}}^2 \sigma _{\text{cor}}^2}}

The final calculations necessary to compute the key rate are the leaked information reconciliation, $\l_{EC}$ and the observed correlations $d(p_{A},p_B)$. The average distance in a typical run for a given sample size can be found by generating appropriately correlated Gaussian distributed strings, binning them and evaluating (\ref{dpe}) directly. For the parameters chosen here, one finds $d(p_A,p_B)\approx 0.1$. For the sample sizes necessary for positive key, the amount of information leaked during reconciliation is well approximated by \cite{Leverrier:2010p150,Furrer:2012p8365},
\eqn{l_{EC} = n_X(H(X_A) - \beta I(X_A:X_B))}
where $H(X) = -\sum_{x\in \mathcal{X}} p(x)\log_2 p(x)$ is the Shannon entropy, $I(X_A:X_B) = H(X_A) - H(X_A|X_B)$ is the mutual information and $0\leq \beta\leq 1$ is the reconciliation efficiency. Recent advances have demonstrated efficiencies as large as 0.94 \cite{Gehring:2015ie}. The probabilities for any given outcome can be found by evaluating the discretised observables in (\ref{quad}) over the appropriate wavefunction. 

\begin{figure}[htb]
\begin{center}
\includegraphics[width=\columnwidth]{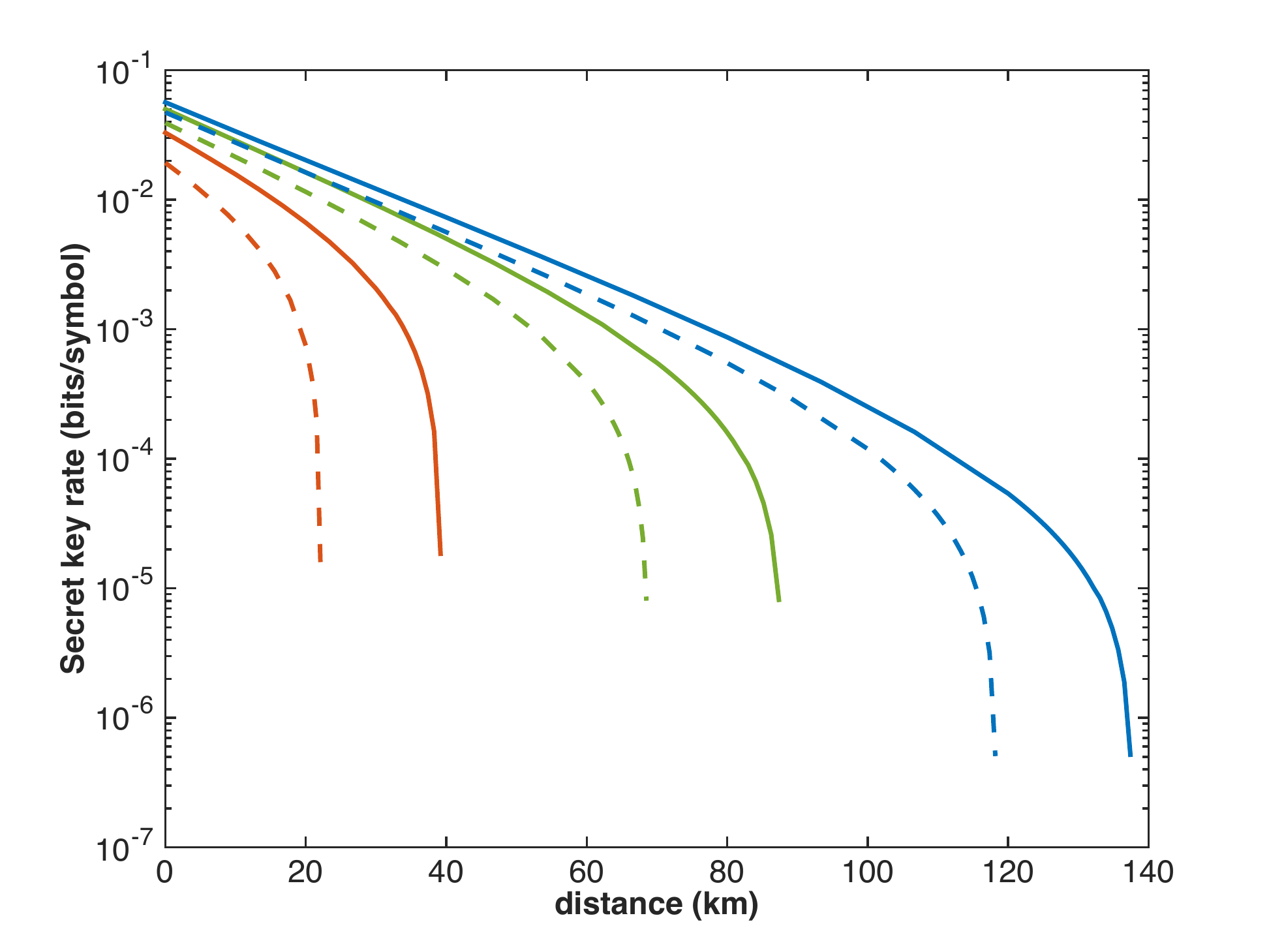}
\caption{Secret key rate as a function of transmission distance for protocols where the key is generated from frequency (dashed) or time (solid) variables. Sample sizes are $N = \{10^9, 10^{10}, 10^{11}\}$ in red, green and blue respectively. Simulation parameters are: $\{\mu_1,\mu_2,\mu_3\} = \{0.2,0.1,0.01\}$, $\{p_{\mu_1},p_{\mu_2},p_{\mu_3}\} = \{0.7,0.2,0.1\}$, $\sigma_\mathrm{coh}$ = 0.5ns, $\sigma_{\mathrm{cor}}$ = 20 ps, $\delta{t}$ = 60 ps, $\delta_\omega = 5$ GHz, $\epsilon = 10^{-10}$, $p_d = 6\times10^{-7}$, $\eta_A = \eta_B = 0.93$, $\beta = 0.94$ and $p_X = 0.5$.} 
\label{rawkey}
\end{center}
\end{figure}

In Fig.~\ref{rawkey} we plot the secret key rate, $l/N$, for various values of the channel uses $N$, as a function to the transmission distance. For the parameters chosen here the protocols where time or frequency are used as the key generating measurement perform comparably. The time-encoded protocol achieves positive key over 40 km for $N = 10^9$ and out to almost 140 km for $N = 10^{11}$. It should be noted that there are many parameters that affect the protocols performance, particularly the source design and decoy state strategy, and a systematic optimisation could further improve performance. In particular, the choice of whether to encode in frequency or time is strongly dependent upon the properties of the source and detectors. For the parameters used here, encoding in the time basis results in higher key rates, but for keeping all other parameters fixed and decreasing the coherence time to $\sigma_{\mathrm{coh}} = 0.3ns$ results in virtually identical rates for both protocols.

\begin{figure}[htb]
\begin{center}
\includegraphics[width=\columnwidth]{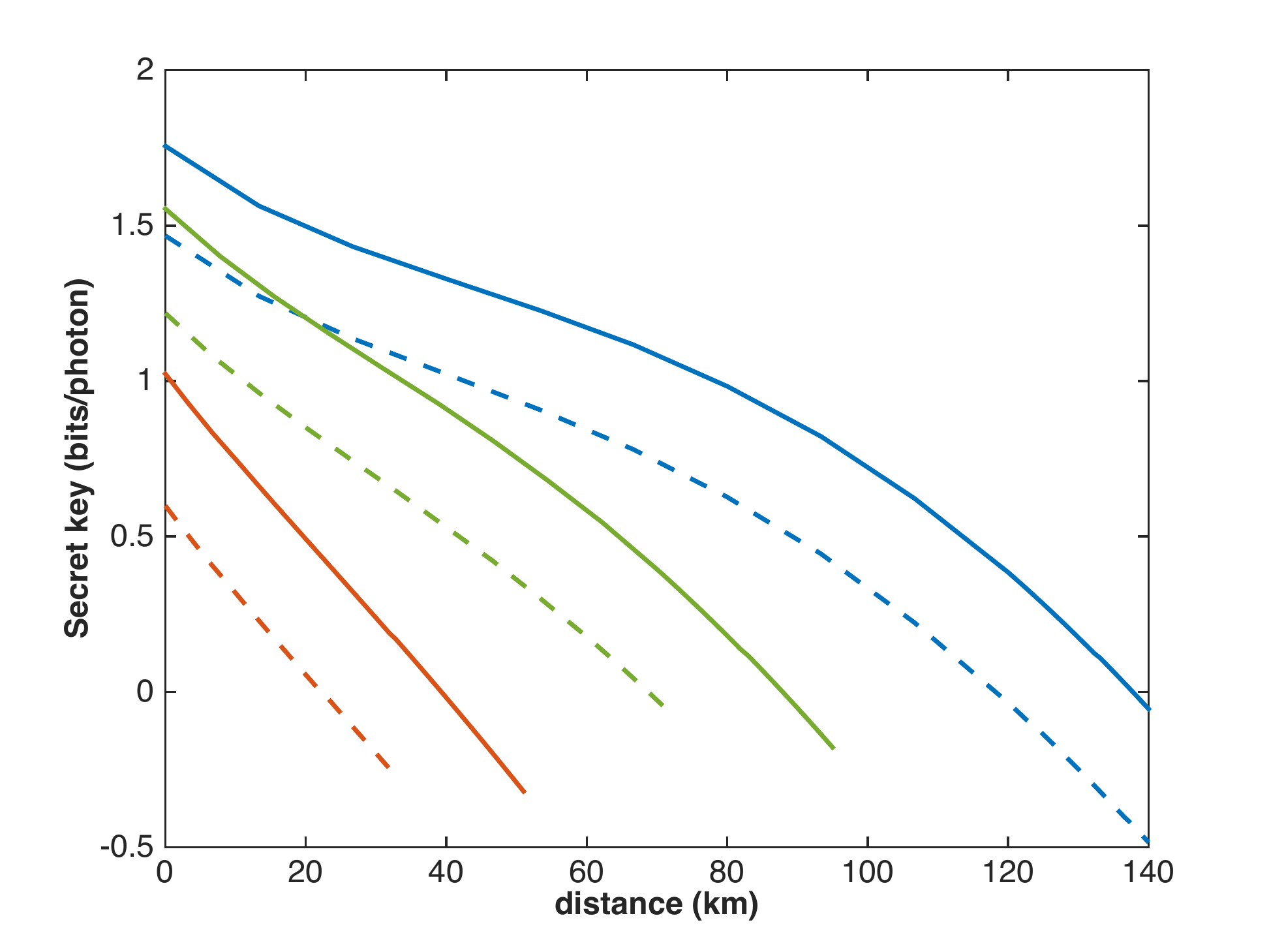}
\caption{Number of secret bits per detected photon as a function of transmission distance for protocols where the key is generated from frequency (crosses) or time (dashed) variables. Parameters the same as Fig.~\ref{rawkey}.} 
\label{pie}
\end{center}
\end{figure}

A second quantity of interest is the photon information efficiency (PIE), the number of secret bits extracted per coincident detection. Recall that one of the attractions of these TFQKD schemes was the promise of a PIE of greater than one bit per photon. In Fig.~\ref{pie} we plot the PIE for the same scenarios as Fig.~\ref{rawkey} and observe a value greater that 1 over distances of ~40 km for $N= 10^10$ and ~90 km for $N=10^{11}$, showing that the protocol is indeed making use of the higher dimensions available.

Arguably the most important quantity however, is the achievable number of secret bits per second. For most protocols this is simply determined by the rate per channel use and the practically achievable clock rate of the relevant source. However, in TFQKD where the key is actually encoded in a temporal variable itself, the question is more involved. In particular, recall we earlier noted that for a positive key we had to ensure the positivity of the statistical fluctuation term $C$. This implies the condition 
\eqn{\epsilon_s/21 > \sqrt{(2g(p_{\Delta_X},n_X))} +\sqrt{(2g(p_{\Delta_P},n_X))}\label{delcon}} which in turn means that both $\Delta_X$ and $\Delta_P$ must be sufficiently large. For the arrival time measurement, the maximum observable value dictates the time frame for a given round, $T_f = 2\Delta_t$ and hence a hard upper limit on the possible clock rate of the protocol of $\frac{1}{T_f}$.

\begin{figure}[h]
\begin{center}
\includegraphics[width=\columnwidth]{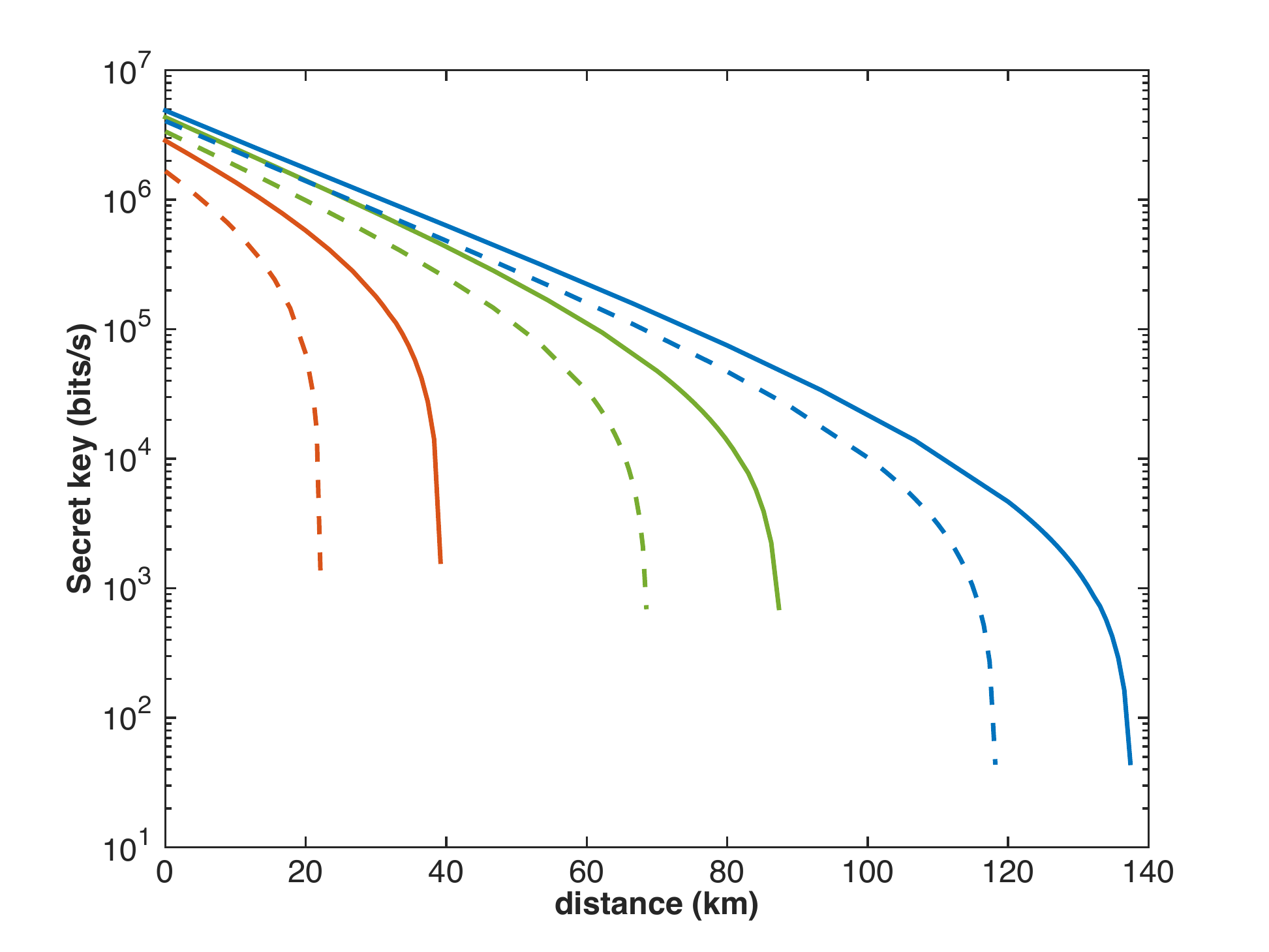}
\caption{Number of secret bits per second as a function of transmission distance for protocols where the key is generated from frequency (dashed) or time (solid) variables. Parameters the same as Fig.~\ref{rawkey}.} 
\label{bps}
\end{center}
\end{figure}

Using our knowledge of Alice's source we can calculate these probabilities for this protocol via Gaussian integration. For a Gaussian distributed variable of variance $V_X$ we have,
\eqn{p_{\Delta_X} = \mathrm{erf}\bk{\frac{\Delta_X}{\sqrt{2 V_X}}}} 
Now for any $\epsilon>0$ if we require $\epsilon >\sqrt{(2g(p_{\Delta_X},n_X))} $ then substituting and rearranging gives,
\eqn{\Delta_X = \sqrt{2 V_X}\mathrm{erf}^{-1}\left[ \bk{1-\frac{\epsilon^2}{2}}^{1/n_X} \right ]}
Applying this to (\ref{delcon}), if we choose to make the two terms on the RHS equal, for the parameters considered here this leads to a requirement on the frequency detection bandwidth of $>290$ GHz or ~5 nm at telecom wavelengths. Similarly we have a requirement on the duration of each round of $T_f>$11.73 ns or a maximum clock rate of 85 MHz. In Fig.~\ref{bps} we plot the number of bits per second assuming the system is run at its maximum clock rate and observe that the system can achieve rates of over a Mb/s up to a distance of 10-20km depending upon the sample size. Furthermore, for $N=10^{11}$ a key rate of ~100 kb/s is possible up to around 90km.

\section{Conclusions}

We have presented a composable security proof for high-dimensional TFQKD, valid against arbitrary attacks and including all finite-size effects. Numerical simulations show that composably secure TFQKD protocols can indeed extract greater than 1 secret bit per detected photon resulting in key rates of over a MB/s at metropolitan distances and maximum range of well over 100km for sufficiently large sample sizes. 

Several avenues for further work remain. Firstly, whilst the proof here has been for the case where Alice and Bob can directly make either spectral or temporal measurements, most concrete proposals for TFQKD involve time-to-frequency \cite{Nunn:2013kf} or frequency to time \cite{Mower:2013tu} conversion. Provided Alice's devices are well characterised it should be straightforward to determine the appropriate uncertainty relation between these effectively conjugate measurements. Secondly there is also a potential weakness to intercept-resend attacks which is particular to TFQKD protocols due to the combination of an in-principle unbounded measurement spectrum and coincidence post-selection as first pointed out in \cite{Nunn:2013kf}. Essentially, the problem is that if Eve makes an extremely precise non-destructive measurement of one observable, say arrival time, this will project onto a state that has limited support within the finite range of Bob's frequency detectors. If Alice and Bob both chose to measure time then Eve will learn this bit and if they both choose to measure frequency, with high probability Bob's detectors will not register the photon and the round will be discarded, opening a loophole in the security. A counter-measure based upon a pre-measurement filtering was proposed in \cite{Nunn:2013kf} which would need to be rigorously incorporated into this proof \cite{prep}. Finally, a remaining unanswered question in all security proofs based upon an uncertainty relation is incorporating an imperfect knowledge of the measurements made by the trusted party. In practice, Alice is not perfectly certain of the POVMs that describe her measurements. A possible solution might incorporate some amount of real time detector tomography into the security analysis.

%Secondly, there is the question of technical practicality. In particular, whilst the detector resolution and bandwidth parameters used in the simulations have either been achieved or are reasonable with present technology, one major idealisation was the assumption that the detectors could operate at the maximum clock rate as determined by the source parameters and chosen failure probability. In practice, current detectors meeting the frequency detection requirements used tend to operate only on the order of 10-100 kHz.

We note that the proof presented here could also be used to rigorously certify the randomness of measurement strings, extending the work of \cite{Vallone:2014ts} to explicitly include a failure probability. This is a particularly attractive possibility since a major strength of these proposals is the high number of bit/photon and hence large overall rates at short distances. 
{\it Note added:} During the writing up of this work the authors became aware of related results by Niu et al. \cite{Niu:2016us}. 
\begin{acknowledgements}
NW would like to thank H.M. Chrzanowski for many helpful discussions. The authors acknowledge funding support from the EPSRC National Quantum Technology Hub in Networked Quantum Information Technologies. J.N. was supported by a Royal Society fellowship.
\end{acknowledgements}

\bibliographystyle{apsrev}

%\bibliography{/Users/nathanwalk/Dropbox/Work/Non-Gaussian-Channel/Latex/paper.bib}

\end{document}